\documentclass{appolb}
\usepackage{graphicx}
\usepackage{dcolumn}
\usepackage{bm}
\usepackage{mathrsfs}
\usepackage{amssymb}
\usepackage{amsmath}
\usepackage{mathtools}
\usepackage[usenames]{color}
\usepackage{relsize}
\def\yR{y_{\raisebox{-0.75pt}{\tiny {\rm R}}}}

\begin{document}

\title{Nuclear Astrophysics in the Multimessenger Era: \\ A Partnership Made in Heaven%
\thanks{Presented at the 2018 Zakopane Conference on Nuclear Physics, Zakopane, Poland}%
}
\author{J. Piekarewicz
\address{Department of Physics; Florida State University; Tallahassee, FL 32306; U.S.A}
}

\maketitle

\begin{abstract}
 On August 17, 2017 the LIGO-Virgo collaboration detected for the first time gravitational
 waves from the binary merger of two neutron stars (GW170817). Unlike the merger of 
 two black holes, the associated electromagnetic radiation was also detected by a host 
 of telescopes operating over a wide range of frequencies---opening a brand new era of 
 multimessenger astronomy. This historical detection is providing fundamental new insights 
 into the astrophysical site for the $r$-process and on the nature of dense matter. In this
 contribution we examine the impact of GW170817 on the equation of state of neutron
 rich matter, particularly on the density dependence of the symmetry energy. Limits on 
 the tidal polarizability extracted from GW170817 seem to suggest that the symmetry 
 energy is soft, thereby excluding models that predict overly large stellar radii. 
 \end{abstract}


  
\section{Introduction}
Almost a century ago in 1915 Albert Einstein published his landmark paper on \emph{``The Field Equations of 
Gravitation''}\,\cite{Einstein:1915}. Shortly after, Einstein predicted the existence of gravitational waves---ripples 
in space-time that travel at the speed of light\,\cite{Einstein:1916,Einstein:1918}. The stretching and squeezing 
of spacetime induces a periodic increase and decrease of the distance between objects that could in principle
be detected by sophisticated laser interferometers. In analogy to electromagnetic waves that are created by 
accelerating charges, gravitational waves are created by time variations of massive objects with a non-uniform 
mass distribution, such as an intrinsic mass quadrupole moment. 

It would take nearly six decades to confirm, albeit indirectly, the existence of gravitational radiation. Back in 1974 
using the Arecibo telescope in Puerto Rico, Hulse and Taylor discovered the first binary pulsar 
(PSR B1913+16)\,\cite{Hulse:1974eb}, a remarkable achievement for which they were awarded the Nobel Prize in 
Physics in 1993. Since initially discovered, the orbit of the binary neutron star system has been slowly and steadily 
shrinking for over 30 years in a manner precisely predicted by the general theory of relativity. At the observed rate 
of energy loss due to the emission of gravitational radiation, the Hulse-Taylor binary pulsar will merge in about 300
million years.

Fittingly, it was near the centennial celebration of the birth of general relativity that the LIGO-Virgo 
scientific collaboration reported the first direct detection of gravitational waves from a binary black hole 
merger\,\cite{Abbott:PRL2016}. On September 14, 2015, shortly after the advanced interferometers were turned
on, a gravitational-wave signal corresponding to a binary black hole merger was detected at both detectors; 
Hanford in Washington State and Livingston in the state of Louisiana. Using theoretical waveforms predicted by
general relativity, individual black holes with masses of about 36 and 29 solar masses merged to produce a 
final black hole with a mass of 62\,$M_{\odot}$. This implies that about 3 solar masses were radiated in the 
form of gravitational waves, or about ten billion times the amount of energy radiate by our own sun in one
year. As impressive, a peak gravitational strain of $10^{-21}$ was measured, suggesting that the 4\,km arms
of both interferometers were stretched and squeezed by a few femtometers. This dramatic discovery opened 
the new and exciting era of gravitational-wave astronomy. 

Soon after, the first detection of gravitational waves from a binary neutron star merger (GW170817) at a 
distance of about 40 Mpc opened the brand new era of multimessenger astronomy\,\cite{Abbott:PRL2017}. 
About two seconds after the arrival of the gravitational-wave signal, the Fermi Gamma-ray Space Telescope 
identified a short duration $\gamma$-ray burst in association with the neutron star merger\,\cite{Goldstein:2017mmi}. 
And within eleven hours of the initial detection, ground- and spaced-based telescopes operating at a variety 
of wavelengths identified the associated \emph{kilonova}---the electromagnetic transient powered by the 
radioactive decay of the heavy elements synthesized in the rapid neutron-capture process ($r$-process).
Distinct features of the kilonova light curve---such as its fast rise, decay, and rapid color evolution from blue 
to red---are consistent with the large opacity typical of the lanthanides, spanning atomic number 57 to 71.
Such characteristic features of the optical spectrum have revealed that about 0.05 solar masses (or nearly 
10,000 earth masses!) of $r$-process elements were synthesized in this single event\,\cite{Drout:2017ijr,
Cowperthwaite:2017dyu,Chornock:2017sdf,Nicholl:2017ahq}. The gravitational wave detection confirmed
the long-held belief of the association of short gamma ray bursts to neutron star mergers. Further, GW170817 
established that binary neutron star mergers play a critical site in the production of heavy elements in the cosmos. 
Finally, constrains on the tidal polarizability (or deformability) of the binary system are starting to provide 
fundamental new insights into the nature of dense matter. Thus, in one clean sweep GW170817 is providing 
compelling answers to two of the ``Eleven science questions for the next century'' \cite{QuarksCosmos:2003} 
identified by the National Academies Committee on the Physics of the Universe: ``What are the new states of 
matter at exceedingly high density and temperature?" and ``How were the elements from iron to uranium made?'' 
Colloquially, one can say that GW170817 created gravitational waves, light, and gold.

The first direct detection of the binary neutron star merger GW170817 is already providing valuable clues into the 
enigmatic nature of dense matter\,\cite{Abbott:PRL2017}. In particular, fundamental properties of the equation of 
state are encoded on the tidal polarizability, an intrinsic neutron-star property that describes its tendency to develop 
a mass quadrupole as a response to the tidal field induced by its companion\,\cite{Damour:1991yw,Flanagan:2007ix}. 
When the separation between the two neutron stars is large relative to their intrinsic size, the gravitational wave profile 
is practically indistinguishable from that of a binary black hole. Yet, as the neutron stars approach each other, tidal 
distortions become progressively more important. Tidal distortions modify the phase of the gravitational wave from its 
point-mass nature and increases the efficiency of gravitational wave emission, thereby precipitating the merger. How 
early during the inspiral phase do tidal effects become important is highly sensitive to the stellar compactness. That 
is, for a neutron star of a given mass, a star with a large radius---and thus a lower average density---is easier to 
tidally distort than a star with a smaller radius and therefore more compact. Essentially, the tidal polarizability probes 
the``fluffiness" of the neutron star. Among the many critical results inferred from GW170817 were relatively small 
tidal polarizabilities that ``disfavor equations of state that predict less compact stars"\,\cite{Abbott:PRL2017}. For
example, for a 1.4\,$M_{\odot}$ neutron star, limits on the tidal polarizability translate into an associated stellar 
radius of $R_{\star}^{1.4}\!\lesssim\!13.76\,{\rm km}$\,\cite{Fattoyev:2017jql,Annala:2017llu}. 

The main goal of this contribution is to examine how the first detection of gravitational waves from GW170817 
improves our knowledge of the equation of state (EOS) of dense matter. Measurement of the tidal polarizability
of the two stars have resulted in valuable limits on the stellar radius of neutron stars, a quantity that has been
traditionally difficult to determine. Indeed, the determination of stellar radii by photometric means has been plagued 
by large systematic uncertainties, often revealing discrepancies as large as 5-6\,km\,\cite{Ozel:2010fw,Steiner:2010fz,
Suleimanov:2010th}. It appears, however, that the situation has improved considerably through a better understanding 
of systematic uncertainties, important theoretical developments, and the implementation of robust statistical
methods\,\cite{Guillot:2013wu,Lattimer:2013hma,Heinke:2014xaa,Guillot:2014lla,Ozel:2015fia,Watts:2016uzu,
Steiner:2017vmg,Nattila:2017wtj}. Nevertheless, that the detection of gravitational waves from a binary neutron 
star merger offers a compelling complement to photometric techniques is a welcome alternative.

We have organized the paper as follows. In Sec.\,\ref{Formalism} we sketch some of the important details
that are needed to compute the tidal polarizability of a neutron star. We then proceed to Sec.\,\ref{Results} 
to display results that highlight the impact of GW170817 on the underlying equation of state. Finally, we conclude 
in Sec.\,\ref{Conclusions}.

\section{Sensitivity of the Tidal Polarizability to the Equation of State}
\label{Formalism}

\subsection{Tolman-Oppenheimer-Volkoff Equations}
The structure of neutron stars is encapsulated in the TOV equations, named after Tolman, Oppenheimer, and 
Volkoff\,\cite{Tol39_PR55,Opp39_PR55}. The TOV equations represent the generalization of Newtonian gravity 
to the domain of general relativity. Remarkably, the only input required for their solution is the equation of state 
of cold (zero temperature) matter in chemical (or ``beta") equilibrium. Indeed, for static, spherically symmetric 
stars in hydrostatic equilibrium the TOV equations may be written as a pair of first order differential equations. 
That is,
\begin{subequations}
 \begin{align}
  & \frac{dP(r)}{dr} = -\frac{G}{c^{2}} \frac{\Big({\mathcal E}(r)+P(r)\Big)
      \left(M(r)+4\pi r^{3}\displaystyle{\frac{P(r)}{c^{2}}}\right)}
      {r^{2}\Big(1-2GM(r)/c^{2}r\Big)} \;, \\
  & \frac{dM(r)}{dr} = 4\pi r^{2}\frac{{\mathcal E}(r)}{c^{2}} \;,
  \end{align}
 \label{TOV}
\end{subequations}
where $M(r)$, ${\mathcal E}(r)$, and $P(r)$ are the mass, energy density, and pressure profiles, respectively. 
Given boundary conditions in terms of a central pressure $P(0)\!=\!P_{c}$ and enclosed mass at the origin 
$M(0)\!=\!0$, the TOV equations may be solved using any suitable numerical solver. Note that the stellar radius 
$R$ and mass $M$ are determined from the following two conditions: $P(R)\!=\!0$ and $M\!=\!M(R)$.  Also note 
that the solution of the problem requires a relation connecting the pressure to the energy density, namely, an 
equation of state.

\subsection{Composition and Equation of State}
\begin{figure}[h]
\begin{center}
\includegraphics[width=0.5\columnwidth]{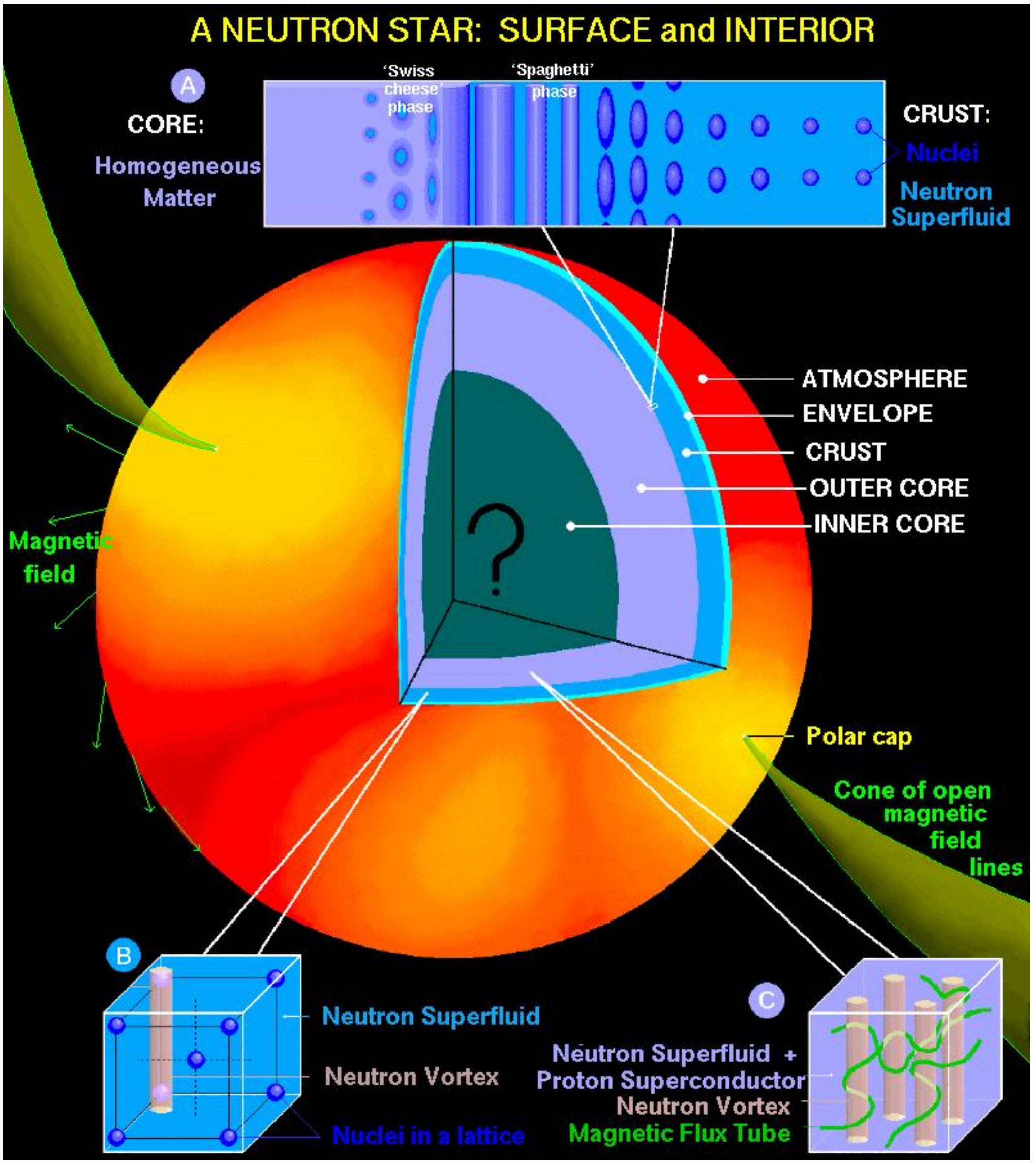}
\caption{An accurate rendition of the structure and phases of a neutron star, courtesy of Dany Page. Of 
              great relevance to the neutron star is the composition of both the stellar crust and core, as well
              as their contribution to the equation of state.}               
 \label{Fig1}
\end{center}
\end{figure}

Although it is fairly well understood how the number of electrons determines the chemistry of the atom and how 
chemistry is responsible for binding atoms into molecules and molecules into both traditional and fascinating new 
materials, one would like to understand \emph{how does matter organize itself} at densities significantly higher 
than those found in everyday materials; say, from $10^{4}\!-\!10^{15}\,{\rm g/cm^{3}}$. In this units the equilibrium 
(or saturation) density of atomic nuclei equals $2.48\times 10^{14}{\rm g/cm^{3}}$. This density, corresponding to
about 0.15 nucleons per cubic fermi, is found in the interior of nuclei. At these enormous densities, it is the pressure 
rather than the temperature that is responsible for squeezing electrons out of the atoms. As depicted in Fig.\,\ref{Fig1},
neutron stars contain a non-uniform crust above a uniform liquid core that is comprised of a uniform assembly of 
neutrons, protons, electrons, and muons in chemical equilibrium. Given that the densities in the stellar core may 
exceed that of normal nuclei by up to an order of magnitude, both electrons and muons contribute to neutralize 
the positive charge carried by the protons. Although the highest densities attained in massive neutron stars is
presently unknown, for soft equations of state---namely, those with a pressure that rises slowly with density---the 
highest density may be such as to favor the formation of new and exotic states of matter\,\cite{Ellis:1995kz,
Pons:2000xf,Weber:2004kj,Alford:1998mk,Alford:2007xm}. However, at densities below saturation density 
other novels phases of matter emerge under these extreme conditions. As the density falls below about 1/2
to 1/3 of saturation density, the average separation between nucleons increases to such an extent that it 
becomes energetically favorable for the system to segregate into regions of normal density (nuclear clusters) 
and regions of low density (a dilute, likely superfluid, neutron vapor).  Such a clustering instability signals the 
transition from the uniform liquid core to the non-uniform crust. The crust itself is divided into an outer and an
inner region. Structurally, the outer crust is comprised of a Coulomb lattice of neutron-rich nuclei embedded in 
a uniform electron gas~\cite{Baym:1971pw,Haensel:1989,Haensel:1993zw,Ruester:2005fm,RocaMaza:2008ja,
RocaMaza:2011pk}. Given that the electronic density increases rapidly with density, it becomes energetically 
favorable for electrons to capture into protons, resulting in nuclear clusters that are progressively more 
neutron rich. Eventually, the neutron excess becomes too large for the nuclear clusters to absorb any more 
neutrons, marking the transition to the inner crust---a region of the star characterized by a Coulomb crystal 
of neutron rich nuclei embedded in a uniform Fermi gas of electrons and a dilute vapor of likely superfluid
neutrons. Even deeper in the crust, distance scales that were well separated in both the crystalline 
phase---where the long-range Coulomb interaction dominates---and in the uniform phase---where the 
short-range strong interaction dominates---become comparable, giving rise to a universal phenomenon 
known as \emph{Coulomb frustration}. Coulomb frustration is characterized by a myriad of complex 
structures radically different in topology yet extremely close in energy---collectively referred to as 
\emph{nuclear pasta} \,\cite{Ravenhall:1983uh,Hashimoto:1984}; see Fig.\,\ref{Fig1}. The fascinating 
and subtle pasta dynamics has been captured using either semi-classical numerical 
simulations\,\cite{Horowitz:2004yf,Horowitz:2004pv,Horowitz:2005zb,Watanabe:2003xu,Watanabe:2004tr,
Watanabe:2009vi,Schneider:2013dwa,Horowitz:2014xca,Caplan:2014gaa} or quantum-mechanical approaches 
in a mean-field approximation\,\cite{Bulgac:2001,Magierski:2001ud,Chamel:2004in,Newton:2009zz,
Schuetrumpf:2015nza}. Yet despite the undeniable progress in understanding the nuclear-pasta phase, no 
theoretical framework exists at present that can simultaneously incorporate quantum-mechanical effects and 
complex dynamical correlations. As a result, a reliable equation of state for the inner crust is still missing. In 
the past, we have adopted a simple polytropic interpolation formula\,\cite{Link:1999ca} to estimate the equation 
of state in the inner crust\,\cite{Carriere:2002bx} and we will continue do so in this contribution.

\begin{figure}[ht]
\begin{center}
\includegraphics[width=0.6\columnwidth]{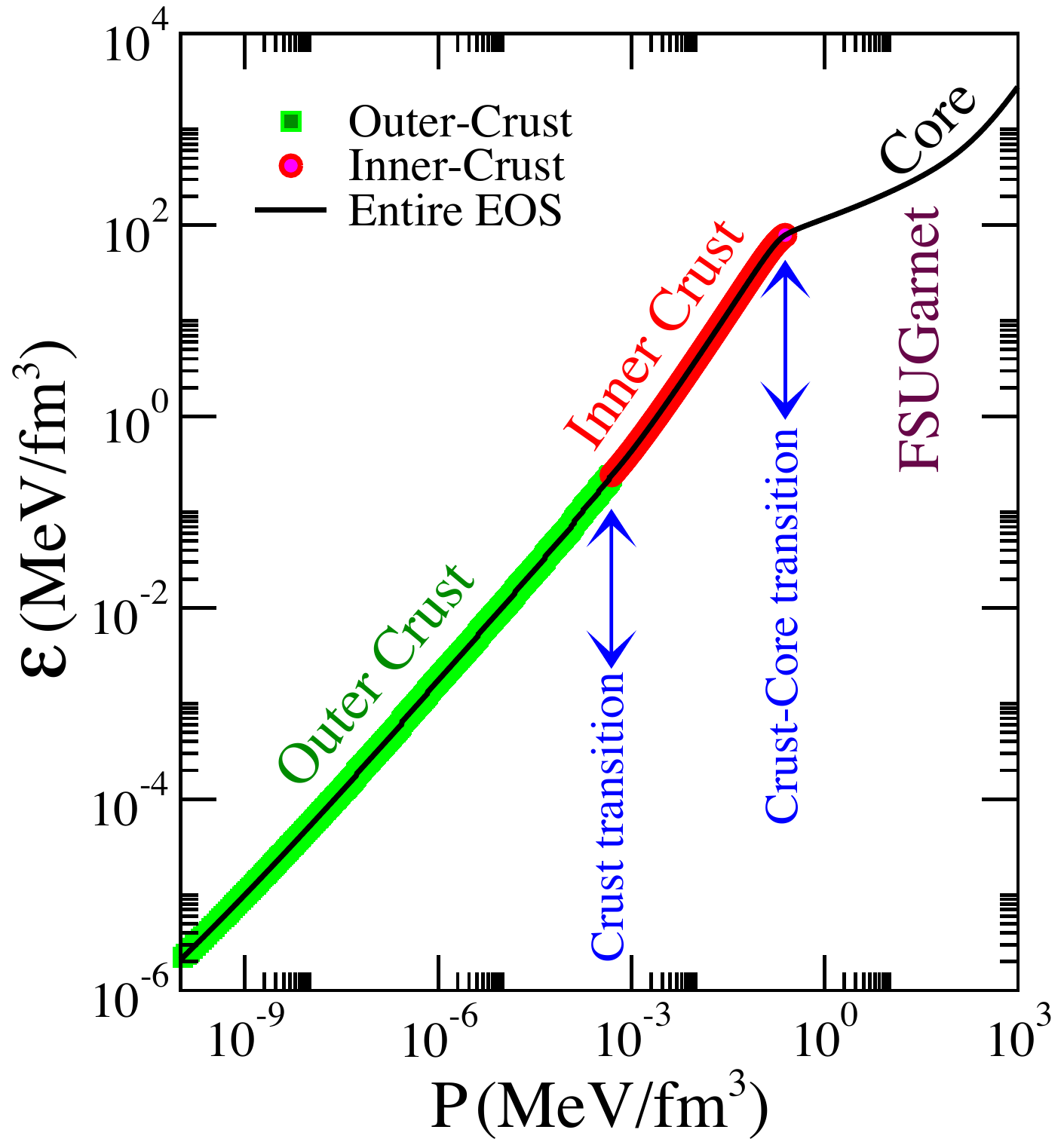}
\caption{Neutron-star-matter equation of state as predicted by the relativistic density functional 
``FSUGarnet"\,\cite{Chen:2014mza}. The composition and equation of state in the non-uniform
crust is described in the text. Note that the features that are mostly sensitive to the choice of 
density functional are the crust-core transition pressure and the EOS in the entire uniform liquid 
core.}             
 \label{Fig2}
\end{center}
\end{figure}
The following prescription is adopted for the neutron-star matter equation of state. First, the EOS 
for the outer crust follows the seminal work of Baym, Pethick, and Sutherland\,(BPS)\,\cite{Baym:1971pw},
slightly modified to incorporate the accurate mass formula of Duflo and Zuker\,{\cite{Duflo:1995}. The 
boundary of the outer crust is determined by demanding that the chemical potential be equal to the bare 
neutron mass. Second, given the complexity of the inner crust, we adopt a polytropic equation of state that 
interpolates between the outer crust and the liquid core\,\cite{Carriere:2002bx}. In particular, the crust-core 
boundary is determined from an RPA analysis that signals the instability of the uniform ground state to cluster 
formation. Finally, The EOS in the uniform liquid core is derived from a relativistic model that is accurate in the 
description of both the properties of finite nuclei and neutron stars\,\cite{Chen:2014sca}. As an example of 
such an equation of state, we display in Fig.\,\ref{Fig2} the predictions from the recently calibrated relativistic 
density functional ``FSUGarnet"\,\cite{Chen:2014mza}.

\subsection{Tidal Polarizability}

Finally, we focus on the tidal polarizability. In the linear regime, {\sl i.e.,} in the limit of weak tidal fields, the ratio 
of the induced mass quadrupole to the external tidal field defines the tidal polarizability. The tidal polarizability is 
the gravitational analog to the electric polarizability of a drop of water. A polar molecule such as water develops 
an electric dipole moment in response to an external electric field. The magnitude of the response is encoded in 
the dielectric constant---an intrinsic property of the material. In the same manner, the response of a neutron star 
to an external tidal field is encoded in the tidal polarizability. Moreover, a time dependent mass quadrupole emits 
gravitational radiation in analogy to the electromagnetic radiation generated by a time dependent electric dipole 
moment. Particularly useful is the \emph{dimensionless} tidal polarizability $\Lambda$ that is defined 
as\,\cite{Abbott:PRL2017}:
\begin{equation}
 \Lambda = \frac{2}{3}k_{2}\left(\frac{c^{2}R}{GM}\right)^{5}
                 =\frac{64}{3}k_{2}\left(\frac{R}{R_{s}}\right)^{5},
 \label{Lambda}
\end{equation}
where $k_{2}$ is the second Love number\,\cite{Binnington:2009bb,Damour:2012yf}, $M$ and $R$ are 
the mass and radius of the neutron star, and $R_{s}\!\equiv\!2GM/c^{2}$ the associated Schwarzschild 
radius. Evidently, $\Lambda$ is a very sensitive quantity of the compactness parameter 
$\xi\!\equiv\!R_{s}/R$\,\cite{Hinderer:2007mb,Hinderer:2009ca,Damour:2009vw,Postnikov:2010yn,
Fattoyev:2012uu,Steiner:2014pda}. In turn, the second Love number $k_{2}$ depends on both $\xi$ 
and $\yR$---a dimensionless parameter that is sensitive to the entire equation of 
state\,\cite{Hinderer:2007mb,Hinderer:2009ca}. The parameter $\yR$ is associated to the non-spherical 
component of the gravitational potential at the surface of the star. In the limit of axial symmetry, the leading 
non-spherical component of the gravitational potential is proportional to the product $H(r)Y_{20}(\theta,\varphi)$, 
where $Y_{20}(\theta,\varphi)$ is the ``quadrupole" spherical harmonic\,\cite{Hinderer:2007mb}. In turn, 
$H(r)$ encodes the dynamical changes to the gravitational potential and satisfies a linear, homogeneous, 
second order differential equation that may be solved in conjunction with the corresponding TOV 
equations\,\cite{Hinderer:2007mb,Hinderer:2009ca,Postnikov:2010yn,Fattoyev:2012uu}. The value of 
$\yR$ is obtained from the logarithmic derivative of $H(r)$ evaluated at the surface of the star and, hence, 
depends on the entire equation of state. Once $\yR$ is known, the second Love number $k_{2}$ can be 
computed and from it---and the compactness $\xi$---the tidal polarizability $\Lambda$.

\section{Results}
\label{Results}
Having defined the entire formalism we are now in a position to display results for the tidal polarizabilities
inferred from GW170817. In Fig.\,\ref{Fig3} we show tidal polarizabilities $\Lambda_{1}$ and $\Lambda_{2}$
associated with the high-mass $M_{1}$ and low-mass $M_{2}$ components of GW170817 using a collection 
of ten relativistic models that provide an accurate description of the properties of finite nuclei and neutron 
stars\,\cite{Fattoyev:2017jql}. The various models differ in their choice of the density dependence of the 
symmetry energy, particularly the slope of the symmetry energy at saturation---a quantity commonly denoted 
by $L$ that is proportional to the pressure of pure neutron matter at saturation. Moreover, the neutron skin 
thickness of ${}^{208}$Pb is a laboratory observable that has been shown to be strongly correlated to 
$L$\,\cite{RocaMaza:2011pm,Brown:2000,Furnstahl:2001un,Centelles:2008vu}. Thus, the various models 
in the figure are labeled using the neutron skin thickness of ${}^{208}$Pb as a proxy for $L$. The combination 
of neutron star masses displayed in the figure are constrained by maintaining the very well measured ``chirp" 
mass fixed at\,\cite{Abbott:PRL2017}:
\begin{equation}
 {\cal M}\!=\!\frac{(M_{1}M_{2})^{3/5}}{(M_{1}\!+\!M_{2})^{1/5}}\!\!=\!1.188\,M_{\odot}.
 \label{Mchirp}
\end{equation}
In turn, the solid circles are used to indicate predictions for a binary system having masses 
of $M_{1}\!=\!1.4\,M_{\odot}$ and $M_{2}\!=\!1.33\,M_{\odot}$, respectively, corresponding to a chirp mass
as in Eq.\,(\ref{Mchirp}). Finally, we show the 90\% probability contour extracted from the low-spin scenario 
assumed in Fig.\,5 of the discovery paper\,\cite{Abbott:PRL2017}. Models to the right of the contour predict 
a symmetry energy that is too stiff, and as a consequence stellar radii that are too large, to be consistent with 
the LIGO-Virgo analysis. 

\begin{figure}[ht]
\begin{center}
 \includegraphics[width=0.6\columnwidth]{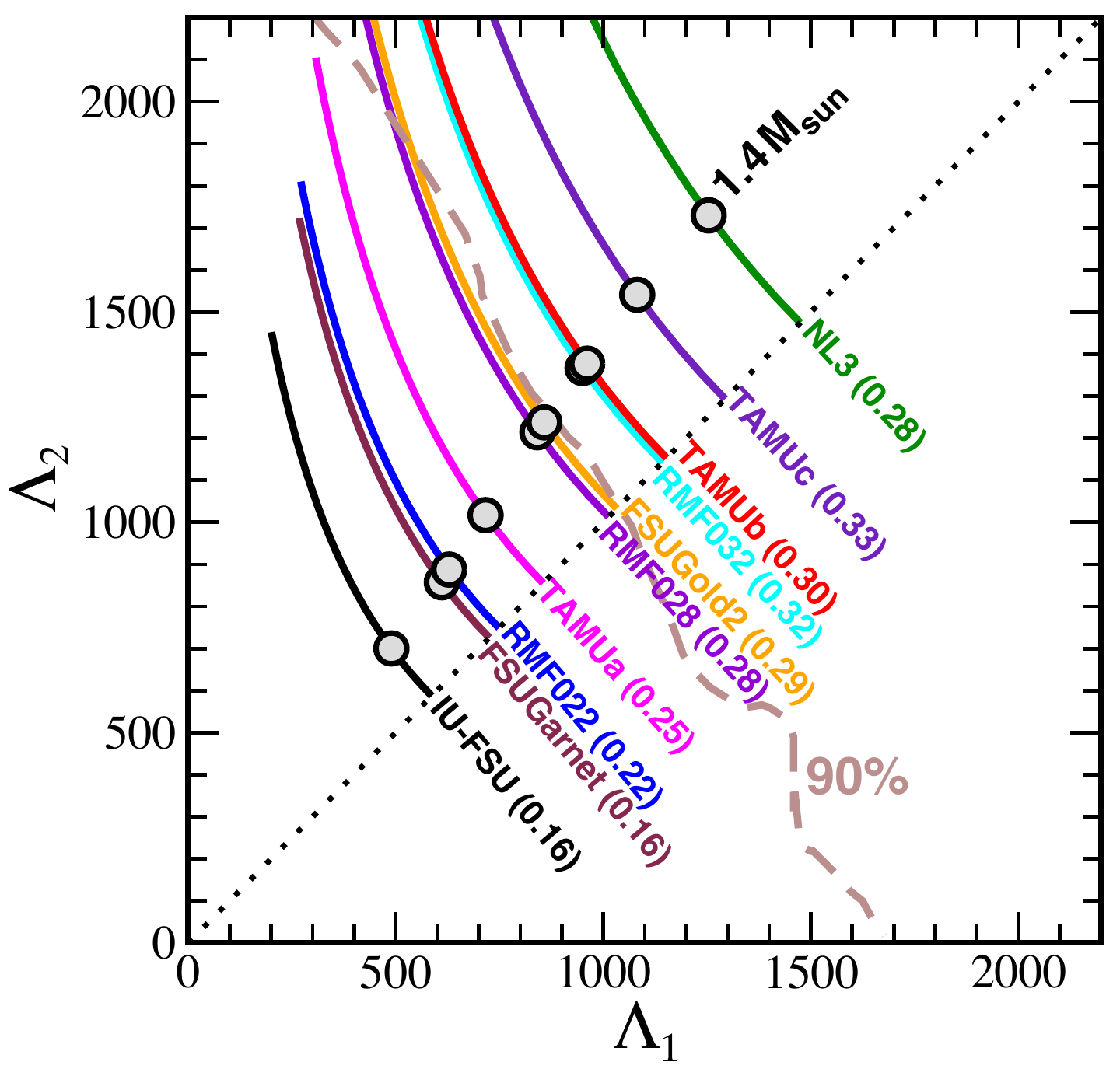}
 \caption{Tidal polarizabilities $\Lambda_{1}$ and $\Lambda_{2}$
    associated with the high-mass $M_{1}$ and low-mass $M_{2}$
    components of the binary neutron star system GW170817 as
    predicted by a set of ten distinct relativistic mean-field 
    models\,\cite{Fattoyev:2017jql}. Models to the right side of the 90\% 
    probability contour extracted from Ref.\,\cite{Abbott:PRL2017} are 
    ruled out. The solid circles represent model predictions for a binary
    system having masses of $M_{1}\!=\!1.4\,M_{\odot}$ and 
    $M_{2}\!=\!1.33\,M_{\odot}$, respectively.}
  \label{Fig3}
 \end{center}
\end{figure}
 
We conclude by displaying in Fig.\,\ref{Fig4} the mass-{\sl vs}-radius relation. For the models of the 
kind described here, the maximum stellar mass is largely controlled by the high density component 
of the EOS of symmetric matter with equal numbers of neutrons and protons. In contrast, stellar 
radii---as well as tidal polarizabilities---are sensitive to the density dependence of the symmetry energy. 
All ten models used in this contribution generate an EOS that is sufficiently stiff to support a 
$M_{\star}\!\approx\!2M_{\odot}$ neutron star\,\cite{Demorest:2010bx,Antoniadis:2013pzd}. However, 
by incorporating the new constraints on the tidal polarizability of a $M_{\star}\!=\!1.4\,M_{\odot}$ 
neutron star, we deduced an upper limit on the stellar radii of 13.76 km\,\cite{Fattoyev:2017jql}. Note 
that a revised analysis of the original GW170817 data that now assumes the same EOS for the two 
stars seems to suggest even more restrictive bounds on the tidal polarizability\,\cite{Abbott:2018exr}. 
Interestingly enough, by combining electromagnetic and gravitational wave information in this new era 
of multimessenger astronomy, additional constraints have been obtained on both the maximum stellar 
mass ($2.17\,M_{\odot}$)\,\cite{Margalit:2017dij} and the minimum radius of a 1.6 solar mass neutron 
star ($10.7$\,km)\,\cite{Bauswein:2017vtn}.
 
\begin{figure}[ht]
\begin{center}
 \includegraphics[width=0.6\columnwidth]{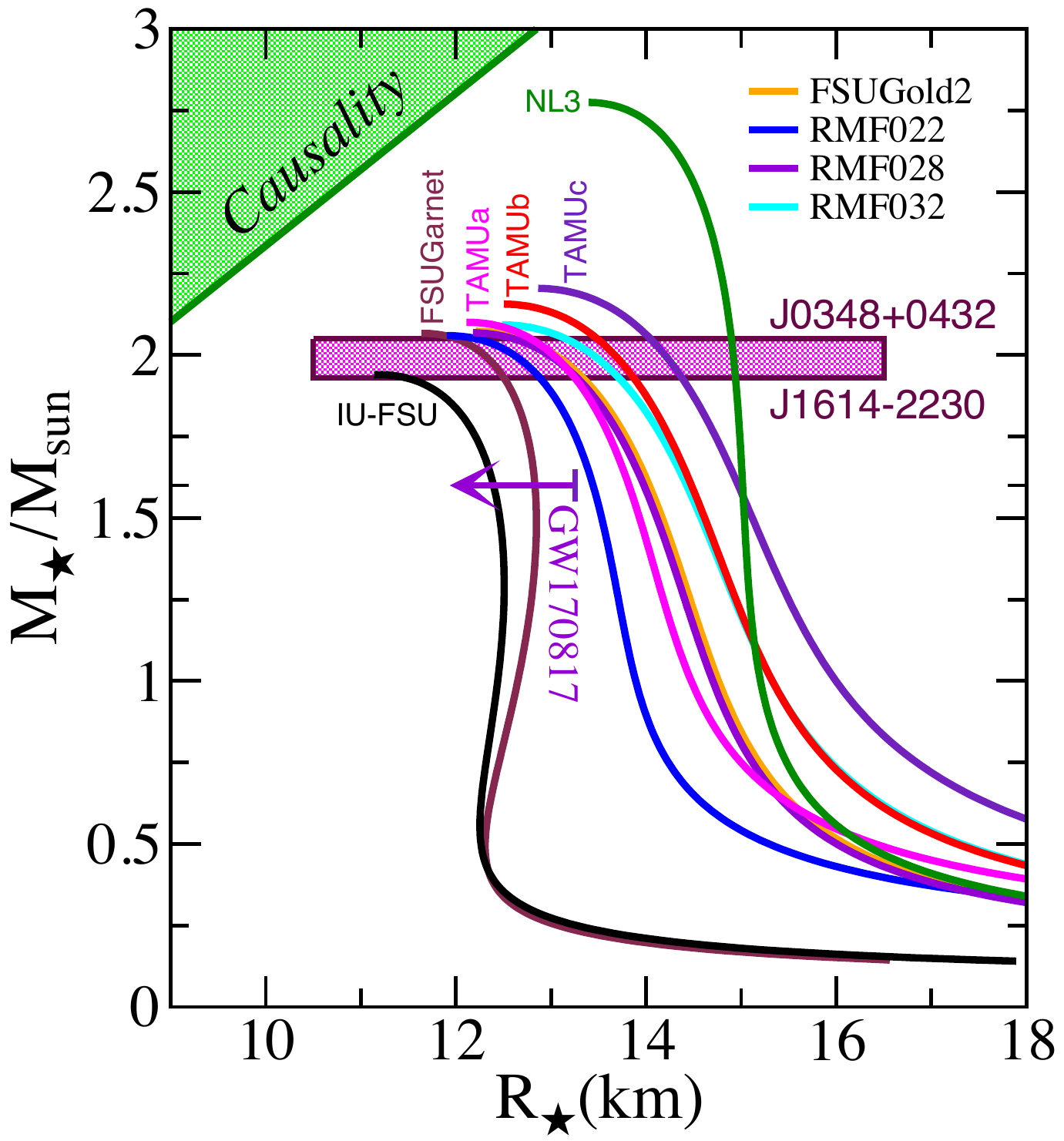}
\caption{Mass-vs-Radius relation predicted by the same ten models used in Fig.\,\ref{Fig3}. 
Mass constraints obtained from electromagnetic observations of two neutron stars are indicated
with a combined uncertainty bar\,\cite{Demorest:2010bx,Antoniadis:2013pzd}.
In contrast, the arrow incorporate constraints on stellar radii obtained exclusively from GW170817 
and exclude many of the otherwise acceptable equations of state\,\cite{Fattoyev:2017jql}. The 
excluded causality region was adopted from  Fig.\,2 of Ref.\,\cite{Lattimer:2006xb}.}  
 \label{Fig4}
 \end{center}
\end{figure}

\section{Conclusions}
\label{Conclusions}

Neutron stars provide a powerful intellectual bridge between Nuclear Physics and  Astrophysics. 
This synergy will strengthen even further with the recent detection of gravitational waves from the
merger of two neutron stars. In this contribution we explored the fascinating structure of neutron 
stars, their connection to nuclear physics through the underlying equation of state, and the new 
limits imposed on the EOS from tidal distortions. The connection between the two fields is strong
because of the sensitivity of the tidal polarizability to the stellar radius, which in turn probes the 
symmetry energy at about twice nuclear matter saturation density. In particular, limits on the 
tidal polarizability of a $1.4\,M_{\odot}$ neutron star translated into an upper limit of 
$R_{\star}^{1.4}\!\lesssim\!13.76\,{\rm km}$ for the associated stellar radius. The multimessenger 
era is in its infancy, yet it is remarkable that the very first observation of a neutron star merger is 
already providing a treasure trove of insights into the nature of dense matter. The third observing 
run by the LIGO-Virgo collaboration is scheduled to start in 2019 and with it the expectation of many 
more detections of neutron star mergers. The future of multimessenger astronomy is very bright
indeed!

\section*{Acknowledgments}
This material is based upon work supported by the U.S. Department of Energy 
Office of Science, Office of Nuclear Physics under Award Number 
DE-FG02-92ER40750.

\bibliography{JPiekarewicz.bbl}
\vfill\eject
\end{document}